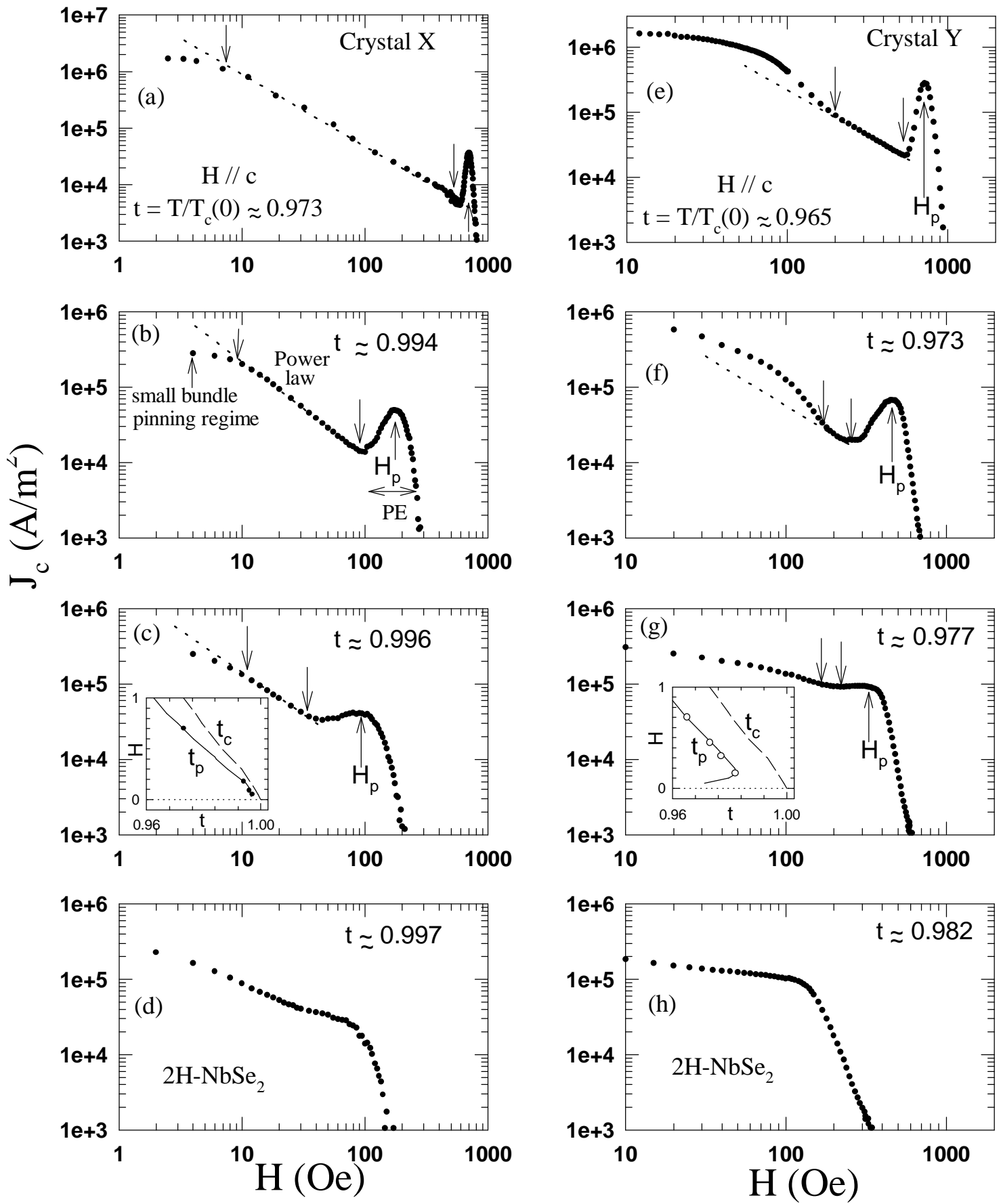

Fig.1(S. S. Banerjee et al)

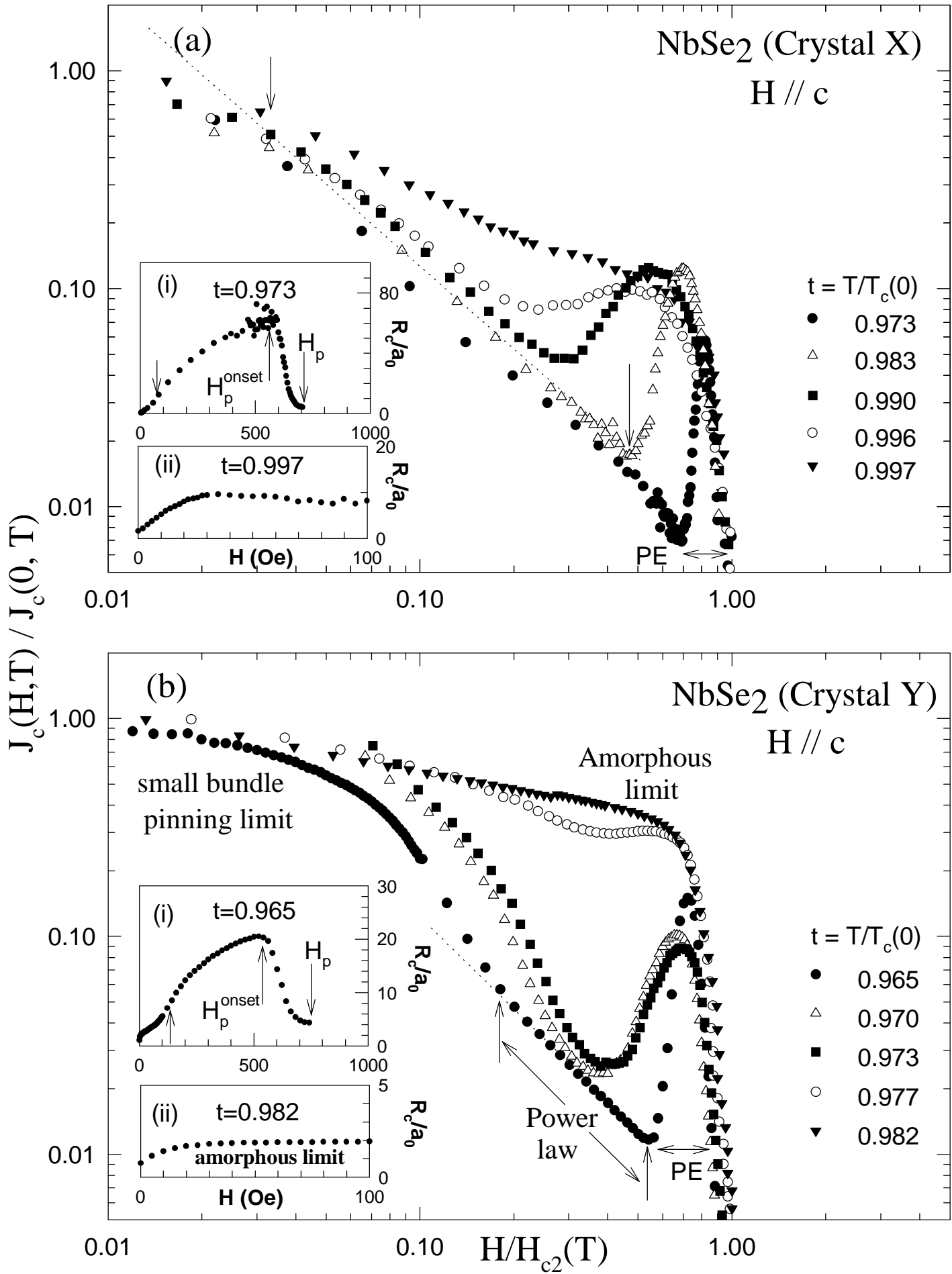

Fig.2 (S. S. Banerjee et al)

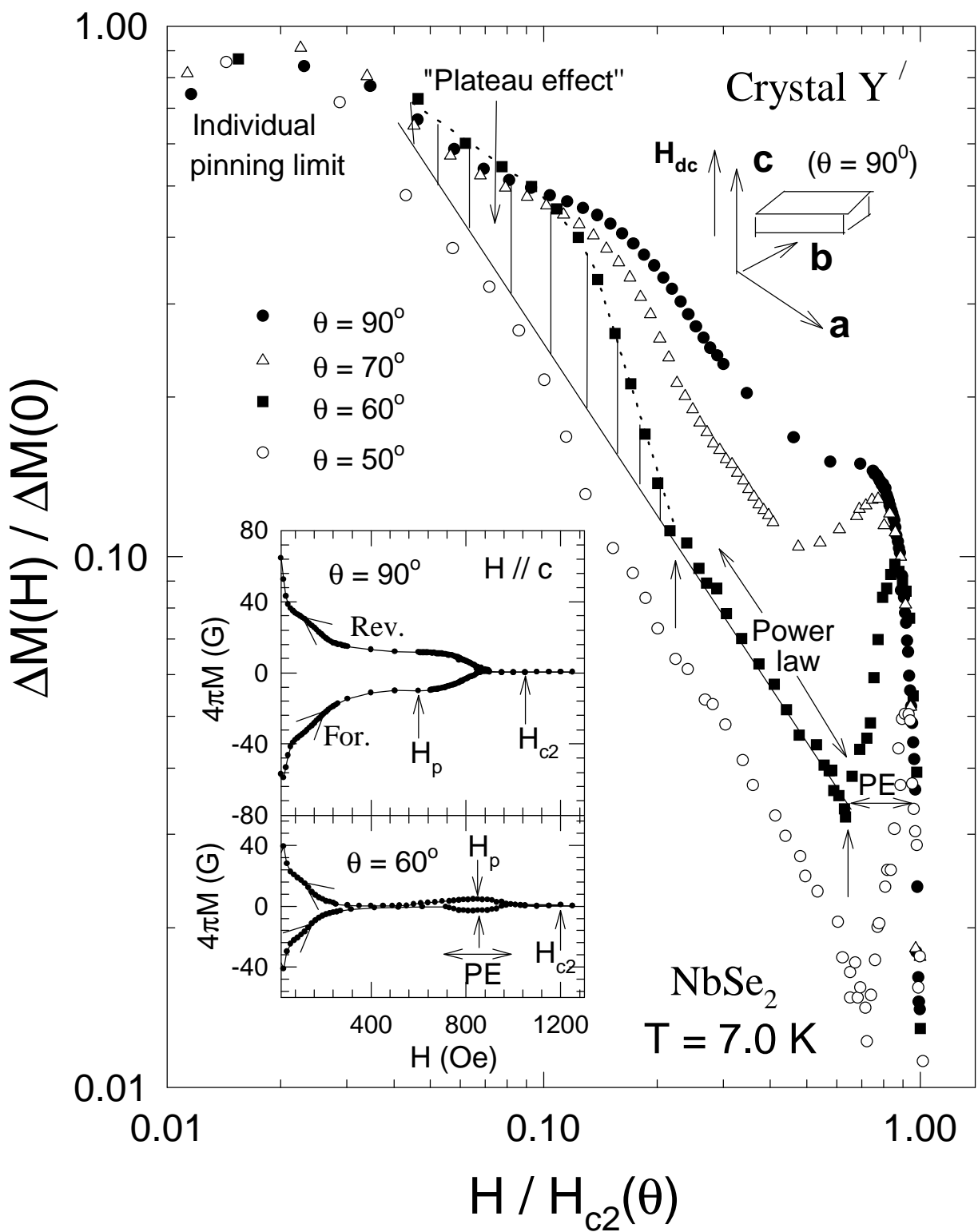

Fig.3 (S. S. Banerjee et al)

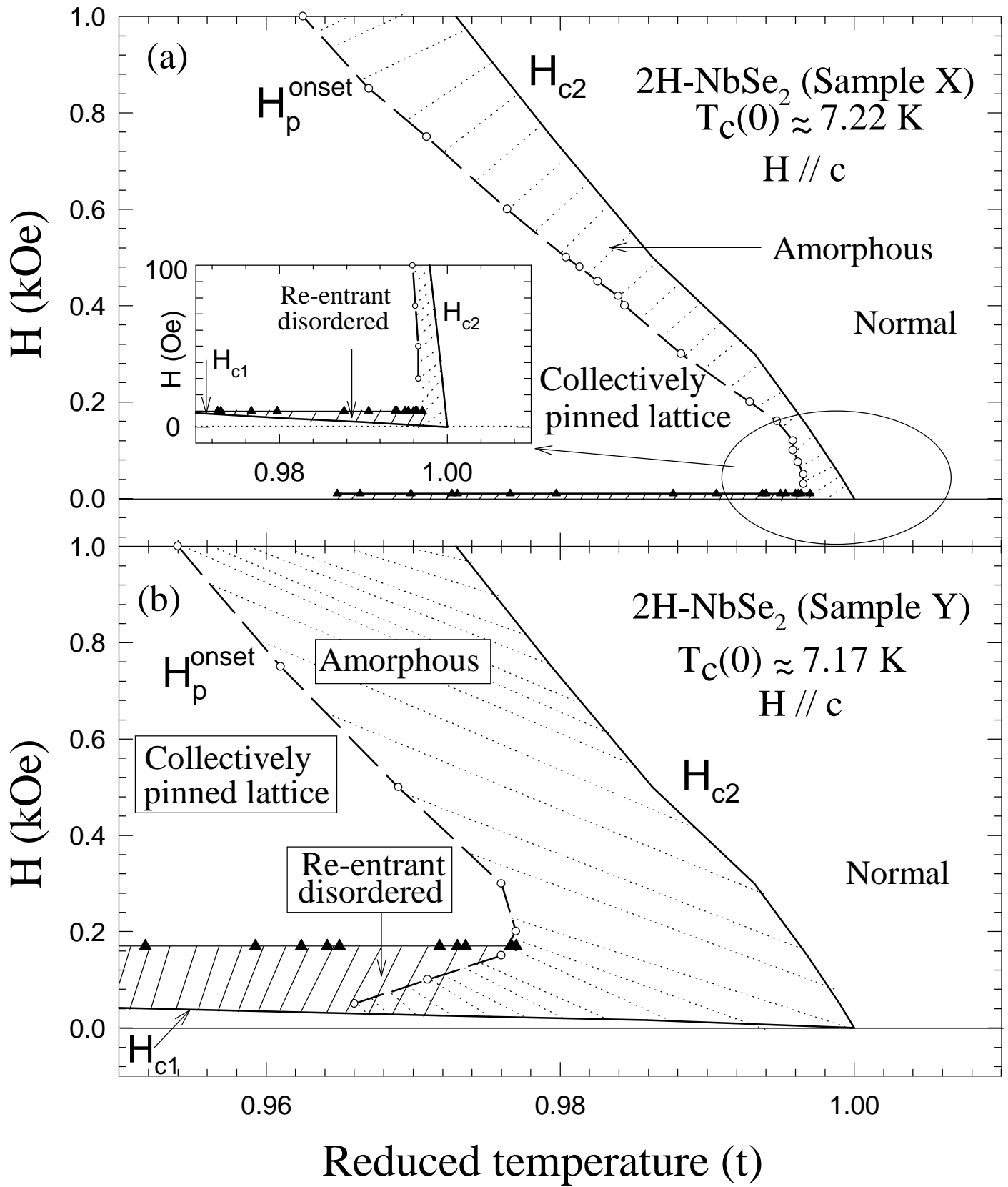

Fig.4 (S. S. Banerjee et al)

# Peak Effect, Fishtail Effect and Plateau Effect : The Reentrant Amorphization of Vortex Matter in 2H-NbSe$_2$

S. S. Banerjee[1,*], S. Ramakrishnan[1] and A. K. Grover[1], G. Ravikumar[2], P. K. Mishra[2] and V. C. Sahni[2], C. V. Tomy[3], G. Balakrishnan[4], D. Mck. Paul[4], P. L. Gammel[5], D. J. Bishop[5] and E. Bucher[5], M. J. Higgins[6] and S. Bhattacharya[1,6,*]

[1] Dept. of Condensed Matter Physics and Material Science, Tata Institute of Fundamental Research, Mumbai-400005, India
[2] TPPED, Bhabha Atomic Research Centre, Mumbai-400085, India
[3] Dept. of Physics, Indian Institute of Technology, Powai, Mumbai- 400076, India.
[4] Department of Physics, University of Warwick, Coventry, CV4 7AL, U.K.
[5] Bell Laboratories, Lucent Technologies, Murray Hill, NJ 07974
[6] NEC Research Institute, 4 Independence Way, Princeton, New Jersey 08540

The magnetic field dependence of the critical current is studied in single crystal samples of the weak pinning type-II superconductor 2H- NbSe$_2$ in the high temperature and the low field region of the (H,T) phase space, in the vicinity of the reentrant peak effect. The experimental results demonstrate various pinning regimes : a collective pinned quasi-ordered solid in the intermediate field that is destabilized in favor of disordered vortex phases in both high fields near H$_{c2}$ and at low fields near H$_{c1}$. The temperature evolution of the pinning behavior demonstrates how the amorphous limit (where the correlation volume is nearly field independent) is approached around the so-called nose region of the reentrant peak effect boundary. Furthermore, the data show that the rapid approach to the amorphous limit naturally yields a peak effect, i.e., a peak in the critical current, in the high field regime, but yields a "plateau effect" instead in the low field regime in an analogous way. With increasing effective disorder the peak effect shifts away from H$_{c2}$ and resembles a "fishtail" anomaly.

## I. INTRODUCTION

The role of quenched disorder and thermal fluctuations in the vortex phases in type-II superconductors is a subject of great current interest [1–5]. Enhanced thermal fluctuations produce a melting transition of the vortex solid into a vortex liquid phase for the high T$_c$ cuprate systems [6,7]. In addition, quenched disorder, i.e., the pinning centers, are expected to add a new variety to the clean system phase diagram in the form of novel glassy phases [2–4,8,9]. The nature of these phases and the transformations amongst them remain a more controversial subject. Furthermore, one expects to find a liquid phase not only at high fields but also at low fields, i.e., a reentrant liquid phase at low induction [10]. Little is known experimentally about this phase. In this context, the observation of a reentrant locus of the peak effect phenomenon in the low T$_c$ superconductor hexagonal 2H-NbSe$_2$ remains a particularly intriguing result [11]. The peak effect phenomenon is the occurrence of an anomalous enhancement of the critical current density, i.e., the pinning force per flux line at high fields near the normal state phase boundary [12] in low T$_c$ systems and nearly coincident with the melting line in the cuprates [13]. The exact causes of the peak effect are uncertain [14–16] but it is widely regarded as the result of a rapid softening of the lattice and the occurrence of plastic deformation [14] and proliferation of topological defects [9,15]. The lattice is expected to be amorphous at and above the peak in J$_c$ [9,17].

Recent theoretical work has focused attention on the possibility of disorder induced glassy phases in the vortex phase diagram [2–4]. Once again much of the attention is focused on the dense phases [8,18,19] and relatively little is experimentally known about the dilute vortex phases [4,20]. A recent theoretical picture [4] proposes that the addition of disorder yields a reentrant glass at low densities, analogous to the low density liquid phase in the disorder free case [10]. This raises the question of how the experimentally observed [11] reentrance of the peak effect in NbSe$_2$ relates to this reentrant liquid [10] or glassy phase [4].

In this paper we focus on the magnetic field dependence of the critical current in the high temperature-low field region of the (H,T) space in the single crystals of 2H-NbSe$_2$. We show explicitly how the pinning evolves with varying H and T in this region from a regime of individual pinning or small bundle pinning to the more collective pinning regime. We track the evolution of the pinning crossovers in samples with different amounts of quenched random disorder and also by exploiting the intrinsic anisotropy of the hexagonal system 2H-NbSe$_2$. These results provide a scenario that mimics the evolution of the characteristics of the critical current density with increasing effective disorder as reported in the cuprates in the low temperature - high field and region [18,19] and as noted in a A-15 superconductor [19] when the quenched disorder was increased externally. The results in 2H-NbSe$_2$ crystals also explicitly show the nature of a *reverse amorphization of the vortex matter* which is actually better described as a *"plateau effect"* that occurs more conspicuously in more disordered samples in the dilute regime. In addition, they also delineate the regime where a collectively pinned ordered vortex phase exists and specifically bring out how an amorphous phase surrounds and/or swamps this ordered regime around the so-called 'nose' of the peak effect boundary [11].

## II. EXPERIMENTAL

We have extracted the field dependence of critical current density J$_c$(H) (for H$\parallel$ c) in two varieties of single crystals of 2H- NbSe$_2$, designated as X [14] and Y [9] respectively, either by directly relating J$_c$(H) to the width of the isothermal magnetization loop or by analyzing the in-phase and out-of-phase ac susceptibility data [21] within the framework of the Critical State Model [22] and following the procedure utilized by Angurel *et al* [23]. The isothermal dc magnetization hysteresis measurements were performed on a Quantum Design SQUID magnetometer with 2 cms. full scan length and/or 4 cms. "half-scan technique" prescribed by Ravikumar *et al* [24], whereas the in-phase and out-of-phase ac susceptibility data with different ac amplitudes were measured using a home built system. The crystal piece X (dimensions 4 × 1.74 × 0.18 mm$^3$) with $T_c(0) \approx$ 7.22 K with resitivity ratio R$_{300K}$/R$_{8K}$ of 20 is similar to the one utilized by Higgins and Bhattacharya [14] in their electrical transport experiments. The sample Y (dimensions 5 × 2 × 0.2 mm$^3$) with $T_c(0) \approx$ 7.17 K and with resistivity ratio R$_{300K}$/R$_{8K}$ of 16 [11] is slightly more strongly pinned than crystal X. However, the locus of peak temperatures T$_p$(H) in this specific sample shows a reentrant characteristic [11,21] below a field value of 150 Oe and near about reduced temperature (t=T/T$_c$(0)) of about 0.98 in the 'nose' region. We have verified that there is a satisfactory agreement between the J$_c$ values (at low fields and close to the 'nose' temperature region) estimated from the width of the dc magnetization hysteresis data and those estimated from an analysis of in-phase and out-of-phase ac susceptibility data [18,19,22]. A simple way to estimate J$_c$ from the in-phase ac susceptibility data [21] is the generalized Critical State Model relationship [22], $\chi' = -1 + \frac{\alpha h_{ac}}{J_c(H)}$, in the limit of full penetration of the ac field h$_{ac}$ into the sample. In this relation, $\alpha$ is a geometrical factor which depends upon the size, shape and orientation of a given specimen w.r.t. applied field H. It can be determined for each circumstance by comparing estimates of J$_c$(H) by different procedures and/or directly measured values of transport J$_c$(H,T). In the field-temperature region of our present interest, J$_c$(H,T) values in crystal X are in the range of $10^4$ – $10^6$ A/m$^2$, whereas those in the crystal Y are about five times larger. The resulting values of the ratio of J$_c$(H,T) to J$_0$(T), the latter being the depairing current density, are in the range $10^{-4}$ to $10^{-3}$, which confirm the weak pinning status of the crystals under investigation.

In addition, we have utilized the anisotropy of the hexagonal 2H- NbSe$_2$ by examining the changes in the characteristics of magnetization hysteresis loops as the applied field is oriented away from the c-axis towards the ab-plane. For such an angular dependence study, we utilized a larger sized crystal (dimensions 5×4×0.45 mm$^3$) with T$_c$(0) $\approx$ 7.25 K. At low fields (H < 200 Oe) and high temperatures, i.e., for 0.96 < T/T$_c$(0) < 1, the locus of $t_p(H)(= T_p(H)/T_c(0))$ values (for H$\parallel$c) in this sample (designated Y$'$) displays behavior similar to that reported in the crystal Y.

## III. RESULTS AND DISCUSSION

### A. Isothermal Critical Current Density for H$\parallel$c

Fig. 1 summarizes the J$_c$ vs H data for H$\parallel$c in the crystals X and Y as two sets of log-log plots in the temperature regions close to the respective T$_c$(0) values. The peaks in J$_c$(H) occur at fields (H$_p$) less than 1 kOe (see insets in Fig.1(c) and Fig.1(g) for $t_p(H)$ curves in X and Y, respectively). We first focus on the shapes of the J$_c$(H) curves (cf. Fig.1(a) to 1(d)) in the cleanest crystal X. In Fig.1(a), the three regimes of J$_c$(H), at a reduced temperature t$\approx$ 0.973, are summarized as follows : (1) At the lowest fields, J$_c$ varies weakly with field, as expected in individual pinning or small bundle pinning regime, noted earlier also by Duarte *et al* [25] and Marchevsky [26]. (2) Above a threshold field value, marked by an arrow, J$_c$(H) variation closely follows the archetypal collective pinning power law [25] dependence ($\sim \frac{1}{H}$). (3) This power law regime terminates at the onset (marked by another arrow) of the peak effect (PE) phenomenon. On increasing the temperature (see Figs. 1(b) and 1(c)), the following trends are immediately apparent : (1) The peak effect becomes progressively shallower, i.e., the ratio of J$_c$(H) at the peak position to that at the onset of PE becomes smaller ; (2) The power law region shrinks and its field dependence weakens. At still higher temperatures (see, for instance, Fig.1(d)), the power law region is nearly invisible and the anomalous PE peak cannot be distinctly identified anymore, only a residual shoulder survives.

In contrast, the second set of plots (see Figs. 1(e) to 1(h)) in the crystal Y show a somewhat different behavior, although the overall evolution in the shapes of J$_c$(H) curves is generically the same. In Fig.1(e), at a reduced temperature t$\approx$ 0.965, one can see the same power law regime as in Fig.1(a), but as the extrapolated dotted line shows, J$_c$(H) departs from the power law behavior in the low field region. The current density in fact, increases rapidly towards the background saturation limit (i.e., the current density at the lowest field end) at significantly higher fields (than those in the crystal X). The smooth crossover to individual or small bundle pinning regime, as seen in the crystal X, therefore adds on an additional characteristic in the crystal Y. Further, with increasing temperature, the power law regime in the crystal Y shrinks faster than that in sample X (cf. Fig.1(e) and Fig.1(f)), leaving only a rather featureless monotonic $J_c(H)$ behavior upto the highest fields (cf. Fig. 1(g) and Fig.1(h)). Note, also, that the limiting value of the reduced temperature upto which the power law regime along with the PE peak survives in the crystal Y is smaller than that in crystal X. This observation reaffirms the notion [11] that the progressive enhancement in quenched disorder shrinks the (H,T) region over which the vortex matter responds like an elastically pinned vortex lattice.

A simple way to understand the evolution of the characteristics in $J_c(H)$ evident in Fig.1 is to recall the conventional collective pinning scenario for weak pinning systems [12], in which $J_c(H)$ could be described in terms of the pinning parameter W and a Larkin volume [27] $V_c$ as,

$$J_c \sim \sqrt{\frac{W(H,T)}{V_c(H,T)}} \qquad (1)$$

The pinning parameter, $W = nf^2$, where $n$ is the density of pins and $f$ represents the strength of the elementary pinning interaction. The Larkin volume $V_c$ is governed by the competition between $W$ and elastic moduli $c_{66}$ and $c_{44}$ of a triangular lattice [12,27]. The field dependence of the elastic moduli is dictated by the reduced field parameter b (=$H/H_{c2}$) at a given temperature. W is commonly described [28] in a separable form as, $W = W_0(T) \times F(b)$, where the function F is described in terms of reduced field b and $W_0(T)$ accounts for the overall temperature dependence of $f$ in a normalized manner. Thus, if we assume that the temperature variation of $W_0(T)$ correlates with the experimental values of current density in zero field ($J_c(0)$), it would be instructive [19] to view the plots of $J_c(H)/J_c(0)$ vs $H/H_{c2}(T)$ at different T.

Figs.2(a) and 2(b) show the resulting plots of $J_c(H)/J_c(0)$ in crystals X and Y of 2H-NbSe$_2$ at some selected temperatures. The evolution of $J_c(H)$ curves in these two sets of plots and its commonality with similar sets of plots in the cases of cuprate superconductor YBa$_2$Cu$_3$O$_{7-\delta}$ [18,19] and the A15 alloy V$_3$Si [19] become now very apparent. Note first, that in X (see Fig. 2(a)), as the temperature increases, the departure from the collective pinning regime, given by power law decay of $J_c(H)$ [25], occurs at smaller values of the reduced field. At the highest temperature (t = 0.997), the conventional sharp peak effect evolves into a broad hump away from the corresponding $H_{c2}$ value (cf. Fig. 2(a) and Fig. 1(d)), somewhat reminiscent of the *Fishtail Effect* (FE). *The evolution of $J_c(H)$ curves from PE to FE in crystals of YBa$_2$Cu$_3$O$_{7-\delta}$ and V$_3$Si has been reported to occur either by progressive increase in quenched random disorder [19] or by progressive decrease in temperature for a given amount of $\delta$ in YBa$_2$Cu$_3$O$_{7-\delta}$ [18,19], in marked contrast to that by the increase in temperature as in the present case of NbSe$_2$.* It is nevertheless reasonable to suggest that the vortex matter becomes amorphous in the field region of the broad hump, and consequently the correlation volume $V_c$ does not vary significantly in this region. Thus, in such a regime (e.g., curves at t=0.997 and t=0.982 in Figs. 2(a) and 2(b), respectively) $J_c(H)$ is expected to track the background behavior in W(H) in eqn.(1). It is pertinent to point out here that at temperatures where PE is very conspicuous, $J_c(H)$ rises from its smallest value in the collective pinning regime at the onset of PE to reach its overall amorphous limit at the peak position (cf. curves at t=0.973 to 0.990 and t=0.965 to 0.973 in Figs. 2(a) and 2(b), respectively), consistent with the framework of the Larkin - Ovchinnikov (LO) [27] collective pinning picture [12,17,23,27]. The $J_c(H)$ curves for the more disordered crystal Y approach the individual pinning limit faster than those in crystal X (compare curves at t=0.965 to 0.982 in Fig.2(b) with those at t=0.973 and 0.983 in Fig. 2(a)).

The above description leads us to propose that at a given temperature, the entire field span is sub-divided into three primary regimes : the single particle or small bundle regime at low fields, a collective pinning of an ordered lattice regime at intermediate fields and finally the departure back to a single particle or amorphous regime at high fields (as marked by arrow at the onset of PE in Fig.2(a)/Fig.2(b)). It is also obvious that the vortex system fails to reach the collective pinning ordered lattice regime at high temperatures very close to $T_c(0)$.

### B. Angular Dependence of Critical Current Density and its Relationship with Evolution in Pinning Behavior

The evolution of pinning crossovers in 2H-NbSe$_2$ system can also be elucidated by examining the changes in the shape of the magnetization hysteresis loop as the direction of the applied field is oriented away from the c-axis of the single crystal.

In such a circumstance, thermal energy remains fixed, but, the field span over which effects of interaction (leading to collective pinning regime) can dominate, expands as a consequence of increase in $H_{c2}$ value following the anisotropic Ginzburg-Landau formalism relationship, $H_{c2} = H_{c2}(\parallel c, T)(sin^2(\theta) + \epsilon^2 cos^2(\theta))^{-1/2}$, where $\epsilon = \frac{H_{c2}(\parallel c)}{H_{c2}(\parallel ab)}$ and $\theta$ is the angle between the applied field H and the ab plane of the single crystal of NbSe$_2$ [1,14]. As $\theta$ changes from $\frac{\pi}{2}$ towards 0, $H_{c2}$ increases from $H_{c2}(\parallel c, T)$ to $H_{c2}(\parallel ab, T)$, and simultaneously the peak field $H_p(\theta)$ also increases as $\frac{H_p(\theta)}{H_{c2}(\theta)}$ remains nearly invariant [14]. For instance, the inset in Fig.3 displays M-H loops at $\theta = \frac{\pi}{2}$ and $\theta = \frac{\pi}{3}$ at T = 7.0 K in crystal Y' of 2H-NbSe$_2$. The main panel of Fig.3 summarizes the angular dependence of normalized magnetization data as log-log plots, following the prescription of Fig.2. It is apparent that in the M-H loop for H ∥ c at 7 K, the collective pinning power law regime sandwiched between the individual pinning limit at low field end and the amorphous limit near the H$_{c2}$ end cannot be distinctly delineated. The M-H loop resembles a *fishtail effect*. However, as the angle $\theta$ reaches $\frac{\pi}{3}$, the three regimes, corresponding to the conventional peak effect near H$_{c2}$, the interaction dominated collective pinning power law region at intermediate fields and the disorder induced rapid approach to the individual pinning limit at low fields (which accounts for the reentrant characteristic in the PE curve), can be easily identified. The solid line and the dashed line drawn for the curve at $\theta = \frac{\pi}{3}$ in the main panel of Fig.3 help to focus on the *reverse* amorphization, i.e., the "plateau effect" as the vortex matter enters the dilute regime (lattice constant a$_0$ of 3500 A$^0$ at H ≈ 200 Oe > penetration depth $\lambda_{ab}$ of 3000 A$^0$ at t=0.97 [29]) from the ordered elastic vortex solid regime. At this juncture, it is tempting to draw an analogy between the Gingras and Huse [4] scenario (of an elastically deformed pinned vortex lattice state sandwiched between higher density vortex glass and very low density "reentrant glass" state) and our experimental observation of a collectively pinned quasi ordered vortex state sandwiched between a highly concentrated amorphous vortex state and a very dilute disordered vortex array in the nearly individual pinning regime. To reiterate, the crossover from collective pinning regime to each of the other two regimes results in an anomalous increase in J$_c$ values. In the case of the upper anomaly, the J$_c$(H) rapidly declines from the peak position of PE due to a collapse of the strength $f$ of individual pins on approaching the H$_{c2}$ value. On the other hand, in the case of the *reverse* anomaly, i.e., the "plateau effect", from the position of the hump, the J$_c$(H) values smoothly cross over to the individual pinning limit.

It is useful to view (see insets (i) and (ii) in both Figs. 2(a) and 2(b)) the computations of radial correlation lengths R$_c$ vs H in crystals X and Y at temperatures corresponding to the two extreme behavior of current density data in the main panel of Figs. 2(a) and 2(b). These computations have been made using J$_c$(H) values in crystals X and Y and following the 2D/3D collective pinning analysis made by Angurel et al [23] as per their Equations (7) and (8). It was surmised by them [23] and we have confirmed [29] by estimating longitudinal correlation length L$_c$ from J$_c$(0) data in the crystal X [14] that the 2D collective pinning description is more appropriate for the crystal X. On the other hand, our estimates of L$_c$ show [29] that 3D collective pinning scenario prevails in crystal Y. The analysis indeed finds (cf. data in the insets in Fig.2(a) and Fig.2(b)) that the values of the ratio R$_c$ / a$_0$ (where a$_0$ is the flux line lattice constant) in crystal X at t=0.983 are larger than those in the more strongly pinned crystal Y at a comparable value of t. The R$_c$ / a$_0$ values in crystals X and Y also appear reasonable in the context of estimates of R$_c$/a$_0$ reported by Angurel et al [23] in much more strongly pinned crystals of 2H-NbSe$_2$. We also note further that the ratio of R$_c$/a$_0$ starts to collapse at H$_p^{onset}$ (see inset (i) in Fig. 2(a) as well as that in Fig. 2(b)), and at the peak field H$_p$, it approaches the amorphous limit as given by its estimate shown in the inset (ii) in Fig. 2(a)/Fig.2(b).

## IV. SUMMARY AND CONCLUSIONS

To summarize, we have presented experimental results on the current density in single crystals of 2H-NbSe$_2$ which reveal the generic nature of the evolution of pinning behavior in weakly pinned vortex matter. The conventional peak effect located near H$_{c2}$, the fishtail effect which spans an extended field region away from H$_{c2}$, the plateau effect located above H$_{c1}$ and the collectively pinned ordered vortex state, are all results of the competition and interplay between interaction amongst vortices and disordering effects caused by thermal fluctuations and quenched random inhomogeneities. In some circumstances, all these effects can be identified as distinct features lying in juxtaposition to each other, whereas in others they are admixed in a manner that the regime of stability of an ordered vortex lattice becomes obscure. The latter kind of behavior has often been reported in recent years in a variety of high temperature cuprate superconductors over a very wide field - temperature span. In the crystals of 2H- NbSe$_2$, such a behavior is seen in a very limited field-temperature region near T$_c$(0) value. It is thus instructive to collate in Fig.4 the domains of a collectively pinned ordered state (cf. power law regimes in Fig.1 and Fig.2) as distinct from the high field conventional PE region and the low field individual pinning limit (and/or the plateau effect) region in crystals X and Y of 2H-NbSe$_2$. In the magnetic phase diagrams shown in Fig.4(a) and Fig.4(b), the field - temperature region between the start of the PE and the H$_{c2}$(T) line has been filled with dotted lines and termed as *amorphous* region, whereas the lower field pinning dominated region has been shaded with solid lines and termed as 'reentrant disordered'. The so called *amorphous* and *reentrant disordered* regions overlap and form a continuum in the neighborhood of the so called 'nose' region of the PE curve (recall t$_p$(H) curves in the insets of Fig.1(c) and 1(g)). Thus, the phase diagrams in Fig.4(a) and Fig.4(b), further clarify how the enhancement in quenched random disorder, as measured by increase in J$_c$ values, shrinks the domain of the collectively pinned and elastically deformed ordered state in the field−temperature region where the interplay between thermal fluctuations and pinning effects predominates. At temperatures, above the nose region, the combination of thermal fluctuations and quenched disorder destabilizes the ordered lattice over the entire field regime.

As a final remark it is worthwhile to recall that Paltiel *et al* [30] have drawn attention to the importance of surface barriers at low fields in crystals of 2H-NbSe$_2$. In the field - temperature region of our present work, the shape of M-H loops (see, for instance insets in Fig.3) suggests that the surface barrier effects are not prominent.

To conclude, we have demonstrated the evolution of the pinning behavior through J$_c$(H) in the high temperature part of the (H,T) phase diagram in weak pinning samples of 2H-NbSe$_2$. The results show a regime of the collectively pinned vortex solid (akin to a Bragg glass [2]) that is destabilized at both high and low fields. At high fields, the critical current increases rapidly to merge with the amorphous limit, and it then decreases rapidly with further increase in H, resulting in a conventional peak effect. At higher temperatures, the peak effect moves away from H$_{c2}$ in the reduced scale and becomes a broader anomaly, strongly resembling a fishtail effect. At low fields, on the other hand, the critical current increases rapidly to merge with the amorphous limit once again, but a weakly field dependent individual pinning regime in the latter case, resulting in what we call a "plateau effect". Both the peak effect and the plateau effect mark an amorphization of vortex matter at the high field and low field limits, respectively. The resulting "phase diagram" of different pinning behavior elucidates the details of the reentrance phenomenon of the peak effect observed in constant H and varying T measurements.

We would like to acknowledge Nandini Trivedi, Satya Majumder and Mahesh Chandran for fruitful discussions. The work at Warwick University is supported by a research grant from EPSRC, U.K.
* e-mail: sb@tifr.res.in / shobo@research.nj.nec.com

FIG. 1. Log-Log plots of J$_c$ vs H for H∥c at selected temperatures in crystals X and Y of 2H-NbSe$_2$. The temperature range for the present experiments was chosen such that the peak fields (H$_p$) were below 1 kOe. The peak effect at H$_p$ has been identified in Figs.1(a) to 1(c) and in Figs.1(e) to Fig.1(g). A pair of downward pointing arrows in each of these figures marks the interval within which J$_c$(H) decay follows a power law dependence expected for collective pinning. At fields above this interval, J$_c$(H) increases anomalously to yield a peak, whereas, on decreasing the field below the lower end of power law region, J$_c$(H) starts to deviate away from the power law behavior to approach the individual pinning limit. The insets in Fig.1(c) and 1(g) display the PE curve $t_p(H)(=T_p(H)/T_c(0))$ and the superconductor-normal phase boundary $t_c(H)(=T_c(H)/T_c(0))$ in crystals X and Y, respectively, as reported in Ref. [11]. The data points lying on the PE curves in each of these insets identify the reduced temperatures at which J$_c$(H) data have been displayed in Figs.1(a) to 1(d) and Figs.1(e) to 1(h). The J$_c$(H) plot at t=0.997 in crystal X in Fig.1(d) and that at t=0.982 in crystal Y in Fig.1(h), show that the peak effect cannot be identified distinctly at the corresponding temperatures. Note the location of these reduced temperature values in the insets of Fig.1(c) and Fig. 1(g); they lie near the "nose" region of the respective $t_p(H)$ curves.

FIG. 2. Log-Log plots of $J_c(H)/J_c(0)$ vs $H/H_{c2}(T)$ for H∥c at selected temperatures in the crystals X and Y of 2H-NbSe$_2$. The PE region has been identified at the lowest reduced temperature for crystals X and Y in Fig.2(a) and Fig.2(b), respectively. The pair of arrows in Fig.2(a) / Fig.2(b) identifies the power law regime. The normalized current density reaches upto a limiting value at the peak. This limiting value is marked as the amorphous limit in Fig.2(b). The insets (i) and (ii) in Fig.2(a) and in Fig.2(b) show $R_c/a_0$ vs H at two reduced temperatures in crystals X and Y, respectively.

FIG. 3. The inset panels show portions of the forward ($-H_{max}$ to $+H_{max}$) and the reverse ($+H_{max}$ to $-H_{max}$) magnetization hysteresis curves at T = 7.0 K at two orientations of the crystal Y′ of NbSe$_2$. The angle '$\theta$' is measured w.r.t. the basal plane of the hexagonal crystal. The PE region, the peak field H$_p$ and the upper critical field H$_{c2}$ are identified in both the hysteresis loops. Note that the H$_{c2}(\theta)$ increases as H orients away from the c-axis towards the basal plane. The main panel displays the normalized values of the width of magnetization hysteresis loop versus H/H$_{c2}(\theta)$ on a log-log plot for various values of $\theta$. Recall that the width of the loop, $4\pi\Delta M(H)=4\pi(M_{rev}-M_{for})$, is a measure of the current density J$_c$(H). Note that the shapes of the curves in the main panel evolve in the same manner as in Fig.2(b). In the curve corresponding to $\theta=60^0$, the two arrows mark the power law regime. The conventional PE region starts at the upper end of the power law regime. The extrapolated solid line passing through the data points in the power law regime and the dotted line passing through the data point in the field region below the lower limit of the power law regime demonstrate the low field anomaly, i.e., "the plateau effect" phenomenon.

FIG. 4. Vortex phase diagrams in the low field - high temperature (i.e., H<1 kOe and $0.95 < T/T_c(0) < 1.0$) region in crystals X and Y of 2H-NbSe$_2$ for fields applied parallel to the c-axis. The region between the onset of conventional peak effect phenomenon ($H_p^{onset}$) and the superconductor-normal phase boundary (H$_{c2}$) has been filled with dotted lines, whereas the low field region below the start of power-law behavior in J$_c$(H) (see Fig.1) has been shaded with solid lines. The filled triangles data points identify the limiting fields above which power law behavior prevails (see, for instance, Fig.1(a) and Fig.1(b)). Note that the collectively pinned quasi-ordered lattice region appears sandwiched between the so-called 'reentrant disordered' region and the amorphous region for $0.97 < t < 0.995$ in crystals X in Fig.4(a) and for $0.96 < t < 0.98$ in crystal Y in Fig.4(b). The curves H$_p^{onset}$ in Figs.4(a) and 4(b) correspond to the temperatures of onset of the peak effect in isofield $\chi'_{ac}(T)$ scans as reported by S. S. Banerjee et al in Ref. [11]. The H$_p^{onset}$ line in Fig.4(b) passing through $T_p^{onset}(H)$ data points in crystal Y taken from Ref.[11] is consistent with the field values (provided H> 200 Oe) at the upper end of collectively pinned power-law behavior which could be determined from the isothermal J$_c$ vs H scans as shown in Fig.1(b). For the sake of completeness, the lower critical field H$_{c1}(T)$ line [11] has also been drawn in each of the phase diagrams.